\documentclass[]{piparticle}
\usepackage{graphicx}
\usepackage{amsmath}

\begin{document}

\runningauthor{E. Goldberg \itshape{et al.}}  

\title{Flow rate of polygonal grains through a bottleneck: Interplay between shape and size}

\author{Ezequiel Goldberg,\cite{inst1}
        C. Manuel Carlevaro,\cite{inst1,inst2,inst4} 
        Luis A. Pugnaloni\cite{inst4,inst3}\thanks{E-mail: luis.pugnaloni@frlp.utn.edu.ar}}
        
\pipabstract{
We report two-dimensional simulations of circular and polygonal grains passing through an aperture at the bottom of a silo. The mass flow rate for regular polygons is lower than for disks as observed by other authors. We show that both the exit velocity of the grains and the packing fraction are lower for polygons, which leads to the reduced flow rate. We point out the importance of the criteria used to define when two objects of different shape are considered to be of the same size. Depending on this criteria, the mass flow rate may vary significantly for some polygons. Moreover, the particle flow rate is non-trivially related to a combination of mass flow rate, particle shape and particle size. For some polygons, the particle flow rate may be lower or higher than that of the corresponding disks depending on the size comparison criteria. 
}

\maketitle

\blfootnote{
\begin{theaffiliation}{99}
   \institution{inst1} Universidad Tecnol\'ogica Nacional - FRBA, UDB F\'isica, Mozart 2300, C1407IVT Buenos Aires, Argentina.
   \institution{inst2} Instituto de F\'isica de L\'iquidos y Sistemas Biol\'ogicos (CONICET La Plata, UNLP), Calle 59 Nro 789, 1900 La Plata, Argentina.
   \institution{inst4} Consejo Nacional de Investigaciones Cient\'ificas y T\'ecnicas, Argentina.
   \institution{inst3} Dpto. Ingenier\'ia Mec\'anica, Facultad Regional La Plata, Universidad Tecnol\'ogica Nacional, Av. 60 Esq. 124, 1900 La Plata, Argentina. 
\end{theaffiliation}
}

\maketitle

\section{Introduction}
\label{intro}

The flowing properties of granular materials are key in a myriad of industrial processes. In this respect, the discharge of a silo or a hopper constitutes the most ubiquitous operation. Considerable attention has been paid to silo discharge in the literature. Since the initial works in the '60 \cite{beverloo1961flow,brown1959} a number of studies have reproduced the essential aspects highlighted in the pioneering works. These are related to the little or no effect played by the height of the granular column (see however a recent experimental report on submerged silos \cite{wilson2014}), and the $5/2$ power dependency of the flow rate with the size of the aperture (see however a study over a wide range of apertures \cite{mankoc2007flow}). The irrelevance of the column height for the flow rate has been recently shown not to be connected with the Janssen's effect \cite{aguirre2010,aguirre2011,perge2012}. Less attention has been put in the material properties and particle shape. In general, material properties seem to have a marginal effect on flow rate \cite{kondic2014,NeddermanBook,brown1970principles}. However, the flow pattern may change from funnel flow to mass flow to 
mixed flow as friction is varied 
either for the particle--particle interaction or the particle--wall interaction \cite{NeddermanBook,brown1970principles}. 

Particle shape is also a factor that has been considered by a number of authors. However, unlike material properties, shape is not defined by a finite number of parameters. For a particular basic shape, for instance rectangles, one can assess the effect of the few parameters that describe that shape (e.g., long side and short side). However, this cannot tell if a different basic shape (e.g., ellipses) will yield different results. As a consequence, the effect of shape has to be surveyed necessarily in a limited manner and the selection of what basic shapes to consider has to be led by the application of interest or by practical issues such as the limitations in fabrication of some shapes or their computer modeling.

Studies on circular grains and on supercircles (rounded squares) in 2D show a significantly lower mass flow rate for the  supercircles \cite{cleary2002}. This authors, compare particles of different shape that have the same larger diameter. We will see that this selection of matching size may bias the conclusions. A number of ulterior works have shown the same overall trend that mass flow rate is reduced in comparison with rounded particles if they present sharp edges \cite{fraigea2008,hohner2012,hohner2013}. A contrasting result was found by some works in which elongated rounded particles flow faster than circles (2D) or spheres (3D); however, they considered frictionless grains in one occasion \cite{langston2004} and in the attempt to validate with experiments they found difficulties in producing particles of same mass and different shape \cite{lia2004}. One particularly interesting exception to the reduced flowability of sharp particles is the case of tetrahedrons that seem to flow faster than spheres \cite{hohner2013}. 
Unfortunately, this study compared spheres and tetrahedrons of different mass/volume, and the conclusion cannot be warranted to be general.

There is a significant point generally overlooked in many of these studies. Since the relative size between the particle and the aperture is the main factor affecting the flow rate, comparison with different shapes has to be done by fixing the particle size. For example, some workers have claimed that hexahedrons and ellipsoids flow faster than spheres based on experiments with different seeds, without paying much attention to the fact that the seeds with non-spherical shape used were also smaller in size \cite{jin2010}. The problem was recognized in other works when comparing tetrahedrons with spheres \cite{hohner2013}. There is an arbitrary decision to make as to what defines if two particles of different shape are of the same size. The canonical solution to this is to define that particles must have the same mass to be considered of the same size. Notice however that some studies use a characteristic length of the particles (e.g., maximum feret diameter) \cite{cleary2002,hohner2012,hohner2013}. Since 
the shape is being changed, keeping a characteristic length constant will lead to different masses. There is no correct selection in this respect. The parameter that defines the effective size has to be chosen according to the particular application, for example. We will show that this choice has a significant impact on the mass flow rate observed in particles of different shapes. 

An additional aspect to the flow of grains is the actual number of particles outpouring per unit time. Although for a given shape this is proportional to the mass flow rate, in comparing results for different shapes the particle flow rate has a different proportionality constant for each shape. Moreover, the criteria used to define if particles of same size are compared will strongly affect the particle flow rate. We will present results that show that to deliver a fixed number of particles per unit time, the use of different particle shapes requires the increase of the aperture in some cases and the reduction in others.

\section{Simulation}
\label{sec:1}

We simulate a 2D silo by means of a discrete element method (DEM) using the Box2D library \cite{box2d}. This package uses
a constraint solver. At each time step of the dynamics, a series of iterations (typically 20) are used to resolve penetrations between bodies through a Lagrange multiplier scheme \cite{catto}. After detecting overlaps, the inelastic collision at each contact is solved and new linear and angular velocities are assigned. The equations of motion are integrated through a symplectic Euler algorithm. The time step $\delta t$ used to integrate the equations of motion is 0.025 $d/g$; with $d$ the diameter of our circular grains and $g$ the acceleration of gravity. Solid friction is also handled by means of a Lagrange multiplier scheme that implements the Coulomb criterion. Box2D is highly efficient when handling complex bodies such as polygons. 

The silo consist of a box $20d$ wide and $200d$ high. Gravity acts in the negative vertical direction. 2000 grains are typically used in each discharge, and these are not reinjected after they leave the silo. After grains initially placed at random without overlaps settle, an aperture opened at the center of the flat base of the silo allows the grains to discharge. For small apertures, clogging may be observed. In such cases the arrested discharge is discarded from the analysis. Only those discharges that did not present clogging are considered. Some material may be left over at the stagnant zones on each side of the orifice. 

\begin{figure}
\includegraphics[width=0.5\textwidth]{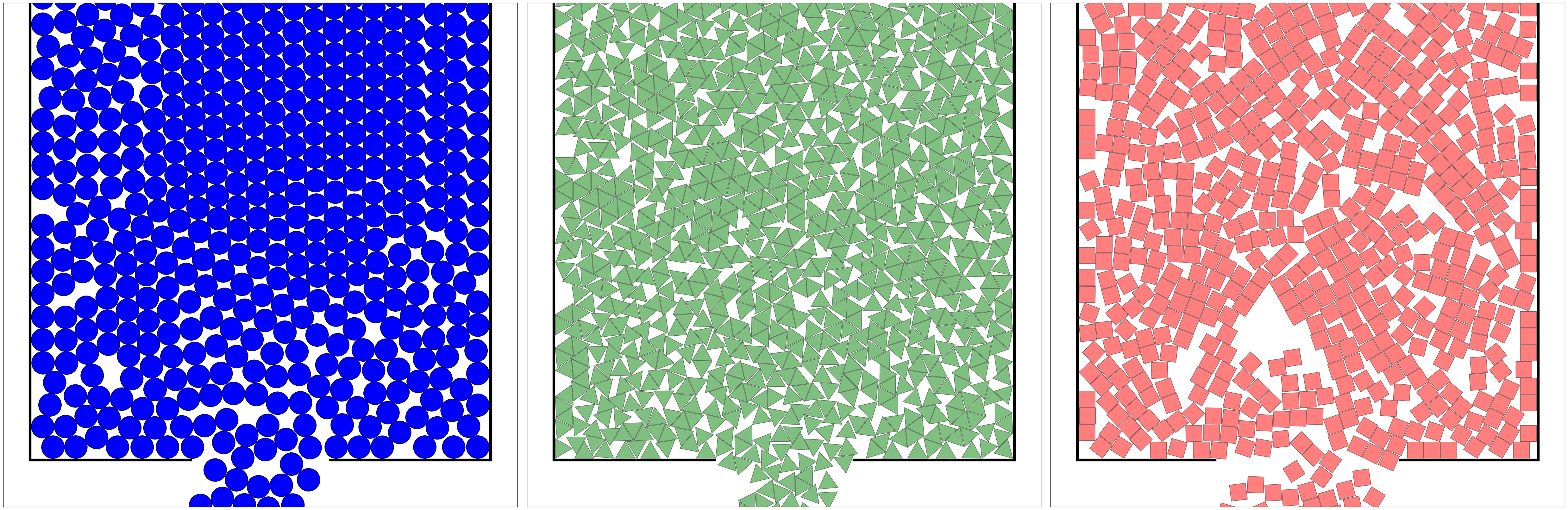}
\caption{Snapshots of the simulations for disks, triangles and squares. Only part of each system is shown.}
\label{fig1}
\end{figure}

Grains of different shapes corresponding to regular polygons (triangles, squares, pentagons, and hexagons) are used in the simulations apart from the circular grains. All grains have the same material density. Each simulation contains only one type of grain and these are monosized. Figure \ref{fig1} shows some snapshots of systems. The restitution coefficient is set in all cases to $\epsilon=0.01$ and the friction coefficient $\mu$ is varied between 0.1 and 2.0. The significantly low restitution coefficient allows for a rapid dissipation of the kinetic energy during the initial filling of the silo. The material density $\rho$ of the grains is set to $0.01$ Kg/m$^2$

After the aperture is opened, we allow for any transient to fade before we calculate the flow rate and other properties. For any simulation, 15 independent realizations with different initial particle positions are carried out to estimate error bars.
 
\section{Results and discussion}
\label{sec:2}

\subsection{Mass flow rate}

\begin{figure}
\includegraphics[width=0.5\textwidth]{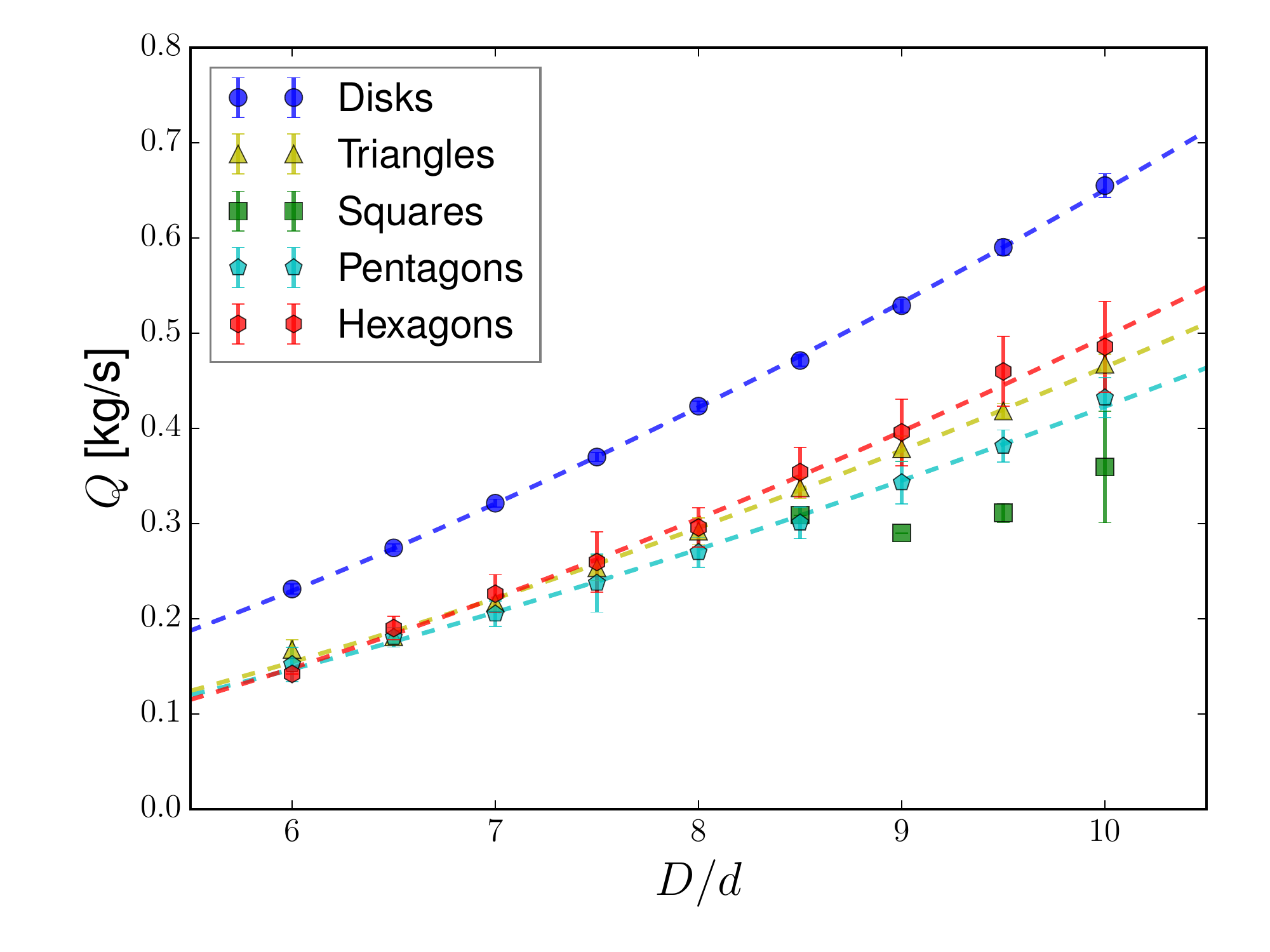}
\caption{Mass flow rate $Q$ as a function of the aperture size $D$ for different particle shapes with $\mu=0.5$. All particle shapes have the same area (mass). The continuous lines are fits to the 3/2-power of $D-kd$ (see fitting parameters in Table I).}
\label{fig2}
\end{figure}

We first consider the mass flow rate $Q$ of grains as a function of the aperture size $D$ for particles of same mass. Figure \ref{fig2} shows $Q$ for different grain shapes. Since simulations are in 2D, the grains have been set with the same material density and area to yield the same mass $m$. Circular grains present values of $Q$ between 30\% and 90\% above those of polygonal grains. Squares and hexagons show larger fluctuation between independent realizations in comparison with other polygons and disks. In particular, squares, which show the lowest flow rate, tend to clog rather often and data on $Q$ for orifices $\leq 8d$ could not be collected. Overall, the flow rate increases in the following order: squares, pentagons, triangles, hexagons and disks.  

We have fitted $Q$ using the 2D Beverloo's rule
\begin{equation}
 Q = C \rho \sqrt{g} (D-kd)^{3/2}. \label{beverloo}
\end{equation}
Where $\rho$ is the material density of the grains, and $C$ and $k$ are fitting parameters. Notice that the bulk density $\rho_b$ is generally used instead of the material density $\rho$ which does not take into account the voidage of the system. In this case, we use $\rho$ and let $C$ to carry this information to avoid taking an arbitrary decision on the region where $\rho_b$ has to be measured. Table I shows the values of $C$ and $k$ obtained for each particle shape and Fig. \ref{fig2} includes the fitting curves. Disks, as well as polygons comply with the 3/2 exponent rule, at least within this narrow range of $D$ explored. The values of $C$ obtained confirm the general trend that the mass flow rate increases with shape in the order described in the previous paragraph. The $k$-value indicates the  intercept with zero flow rate and constitute an extrapolated estimate of the minimum size of the orifice necessary to observe flow. In practice, this estimate obtained with data at relatively large values of $D$ is little reliable. It has been shown that a more careful analysis of the flow at smaller $D$ where intermittent flow can still be analyzed, suggest that a different equation to that of Beverloo is more appropriate by considering that only at $k=1$ would $Q$ become strictly  zero \cite{mankoc2007flow}. Nevertheless, the value of $k$ gives an indication of how soon the flow rate drops as $D$ is decreased. For disks, we find $k\approx 2.0$, which is consistent with experiments where a similar range of $D$ were used \cite{aguirre2010} (if smaller $D$ are considered $k$ may drop down to $\approx 1$ \cite{mankoc2007flow,kondic2014}). Table I indicates that disks and pentagons are predicted to stop flowing at $D\approx 2 d$. However, triangles and hexagons would only flow for somewhat larger $D$. Again, these are only rough indicators, since measurements at lower values of $D$ may lead to different estimates, particularly if states of intermittent flow are included in the analysis.

These results show that shape matters in granular flow as shown by others \cite{cleary2002,fraigea2008,hohner2012,hohner2013,langston2004,lia2004}. However, the variability of $Q$ between different faceted grains is rather limited in comparison with the significantly higher mass flow rate of disks. In general, setting apart triangles, the more vertices a polygon has, the higher $Q$ is. This is consistent with the idea that polygons with many facets should tend to the behavior of circular grains.

One may speculate that the use of faceted particles changes particularly the frictional character of the interactions. While disks may roll on their contacts, polygons will mostly slide side-to-side on each other. A polygon may be forced to ``roll'' (rather than slide) on top of flat inclined surface if the friction angle $\theta$ (i.e., $\theta=\tan \mu$) exceeds the maximum angle of stability $\theta_{stab}$ (defined as the angle of inclination of a flat supporting base at which the vertical line that passes through the center of mass of the polygon meets one of its vertexes). Hence, it is expected that changing the friction coefficient may lead to different effects on the polygons dynamics and hence on the flow rate. The corresponding values of $\mu$ beyond which a polygon would roll before sliding are $\mu>\sqrt{3}\approx 1.732$ for triangles, $\mu>1$ for squares, and $\mu>\sqrt{5+2\sqrt{5}}\approx 0.727$ for pentagons.

\begin{figure}
\includegraphics[width=0.5\textwidth]{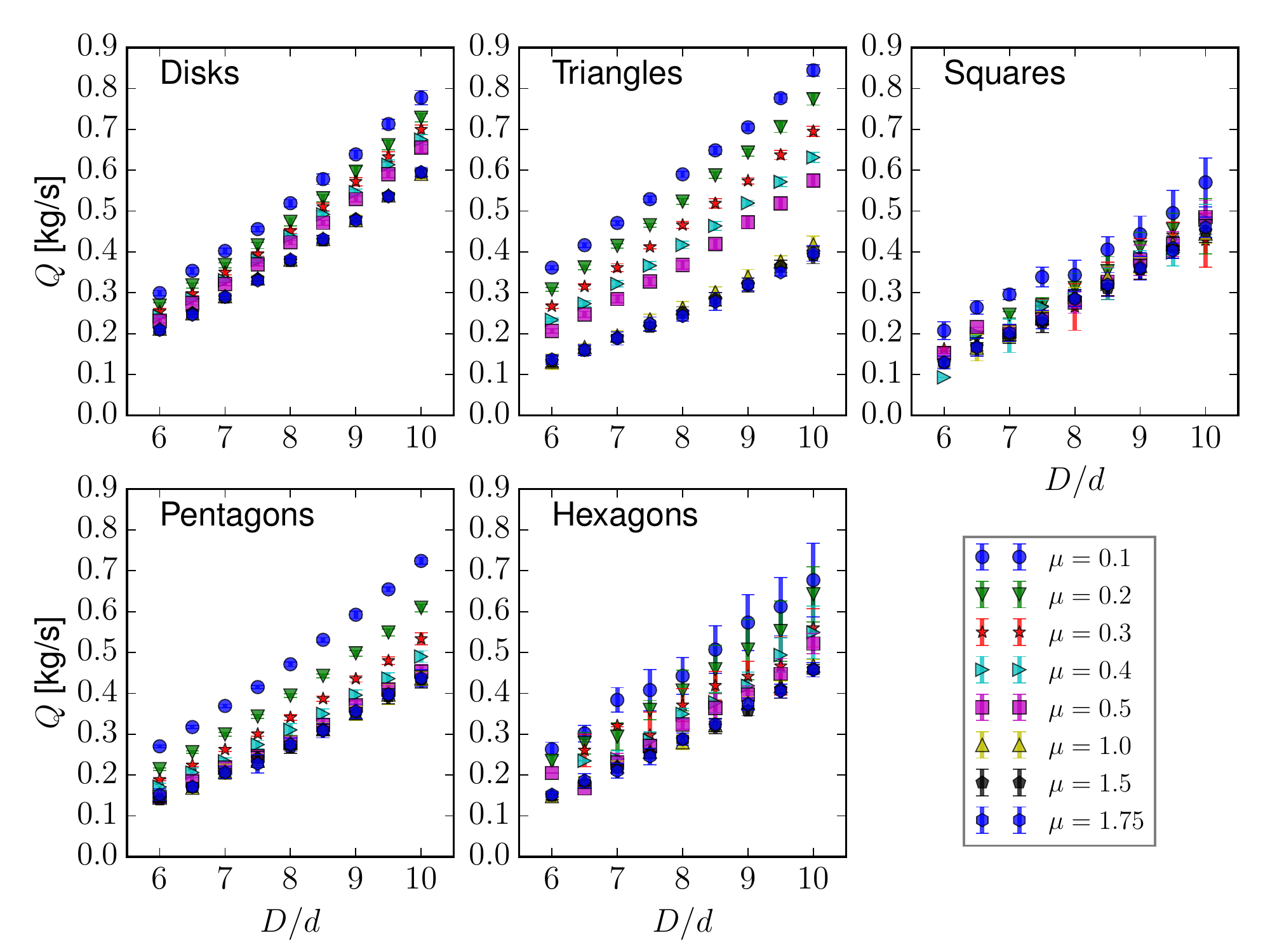}
\caption{Mass flow rate as a function of $D$ for different friction coefficients for each particle shape: disks, triangles, squares, pentagons and hexagons. All particle have the same area (mass) irrespective of their shape.}
\label{fig3}
\end{figure}

Figure \ref{fig3} shows $Q$ for different $\mu$ for each grain shape. As we can see, friction does not induce any significant change in flow rate if $\mu>1.0$. However, a considerable increase in $Q$ is observed if we lower $\mu$ below 0.5. The one exception is found in squares that are little affected by any change in $\mu$. The increase in $Q$ with decreasing $\mu$ is consistent with previous simulations of disks \cite{kondic2014}. Interestingly, there are not clear evidences of a qualitative change when $\mu$ is higher than the maximum stability angle for each polygon. We presume that the interaction with other touching grains is as important as the interaction with any eventual grain that serves as primary support for a given polygon. Since the stability angle defined is only based on the sole interaction with a base, the expectation that a qualitative feature may be observed at these $\mu$ values may be unjustified due to the complex multiparticle interactions.

\begin{table} \label{tabla1}
\caption{Fitting parameters for the Beverloo equation in 2D as defined in Eq. (\ref{beverloo}). For squares, the poor statistics due to clogging events prevent us from extracting reliable fits.}
\begin{center}

 \begin{tabular}{l r r }
 \hline\hline\\
 Shape & C & k \\
 \hline\\
  Disks & 0.92 &  2.01\\
  Triangles & 0.70 & 2.30\\
  Squares & -- & --\\
  Pentagons & 0.61 & 2.09\\
 Hexagons & 0.82 & 2.77\\
 \hline
 \end{tabular}
 
\end{center}

\end{table}

\subsection{Velocity, density and energy profiles}

The reduced flow rate of polygonal grains may be caused by a drop in the packing fraction $\phi$ and/or a fall in the actual velocity of the grains. The packing fraction of polygonal grains in a rectangular box subjected to tapping has been previously considered \cite{carlevaro2011}. It was shown that regular polygons able to tessellate the plane (triangles, squares and hexagons) in general arrange in higher packing fractions as it may be expected. Pentagons, on the other hand, yield lower packing fractions in comparison with monosized disks. Despite this, the packing fraction during flow does not need to be connected with the values found in static systems. Flow implies shear, which induces Reynolds dilatancy, which in turn reduces $\phi$ if the initial compaction is above the dilatancy onset.

\begin{figure}
\includegraphics[width=1.0\columnwidth]{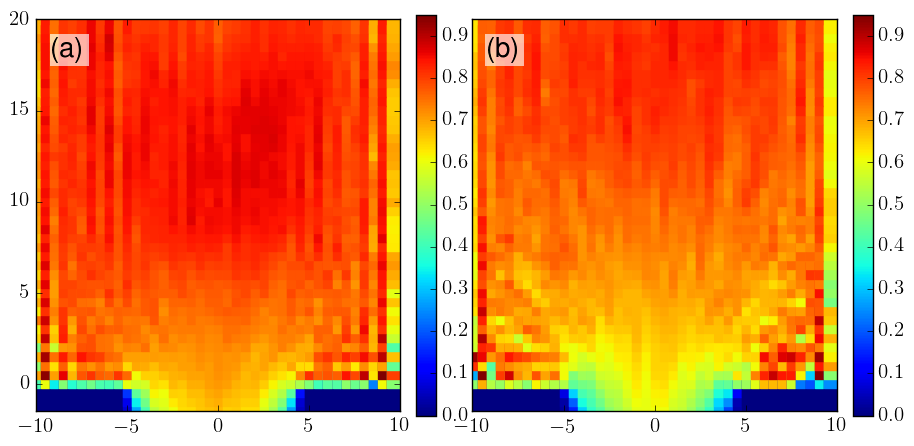}
\caption{Mean packing fraction distribution for disks (a) and triangles (b) averaged over a discharge. The aperture is $D=10d$ and the friction coefficient $\mu=0.5$. Disks and triangles have the same area (mass).}
\label{fig4}
\end{figure}

We have measured $\phi$ profiles across the silo during a discharge for the case of disks and triangles of the same mass (see Fig. \ref{fig4}). The calculations are done by using the  package Shapely for polygon intersections \cite{Shapely}. The time average of $\phi$ during the discharge measured in cells of $1d \times 1d$ are considered. As we can see, $\phi$ is lower for triangles than it is for disks in general. This is in contrast with the higher packing fractions found for triangles in tapped static systems \cite{carlevaro2011}. It is clear that the reduction in $\phi$ observed in the area of the outlet stretches farther into the silo for triangles. At the aperture, $\phi$ is smaller overall for triangles and the effective ``empty annulus'' (i.e., the depleted region next to the edges of the aperture) looks thicker for triangles too. This explains the high values of $k$ found in the fits of the Beverloo rule for triangles (see Table I), which is consistent with experimental results on sharp grains like some sand samples \cite{beverloo1961flow}. The reduced mass flow rate observed for polygons may be then caused by the reduction in $\phi$. However, the 
reduction with respect to circular particles is less than 15\%, which is insufficient to explain the at least 30\% change in $Q$. Hence, one should expect that also the mean velocity of the exiting grains should be higher for disks.

\begin{figure}
\includegraphics[width=1.0\columnwidth]{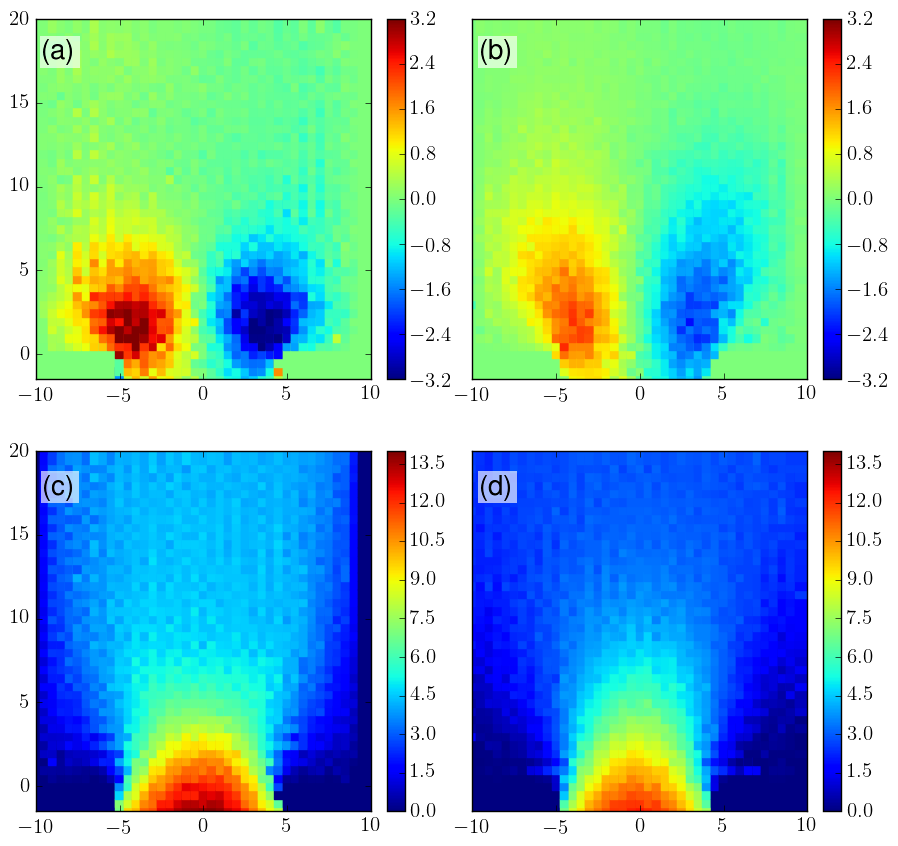}
\caption{Mean particle velocity: (a) $v_x$ for disks, (b) $v_x$ for triangles, (c) $v_y$ for disks and (d) $v_y$ for triangles. The aperture is $D=10d$ and the friction coefficient $\mu=0.5$. Disks and triangles have the same area (mass). The color scale indicates the velocity in ms$^{-1}$.}
\label{fig5}
\end{figure}

Figure \ref{fig5} shows the profiles for the vertical and horizontal velocity of disks and triangles. The mean horizontal velocity is different from zero only close the aperture in a V-shaped region consistent with convergent streamlines towards the outlet. Triangles show a more stretched region of non-zero horizontal velocity in agreement with the stretched region of low $\phi$ [see Fig. \ref{fig4}(b)]. Disks present two hotspots above each edge of the outlet with high horizontal mobility in clear contrast with the triangles that present a smoother profile. From the mean vertical velocity one can observe clearly a more significant stagnant zone for triangles on each side of the aperture. This is consistent with results of several authors (see for example, \cite{cleary2002,hohner2013}) and with the intuition that sharp edged grains will display a higher effective angle of repose. This taller stagnant zone seems to be responsible for inducing shear on a larger area for triangles, leading to the reduced $\phi$. 
Also the vertical velocity at the aperture is in average higher for disks; roughly 9\%. The combined effect of reduced $\phi$ and reduced vertical velocity adds to a $\approx 25\%$ change in $Q$ when comparing disks and triangles at this opening width.   

\begin{figure}
\includegraphics[width=1.0\columnwidth]{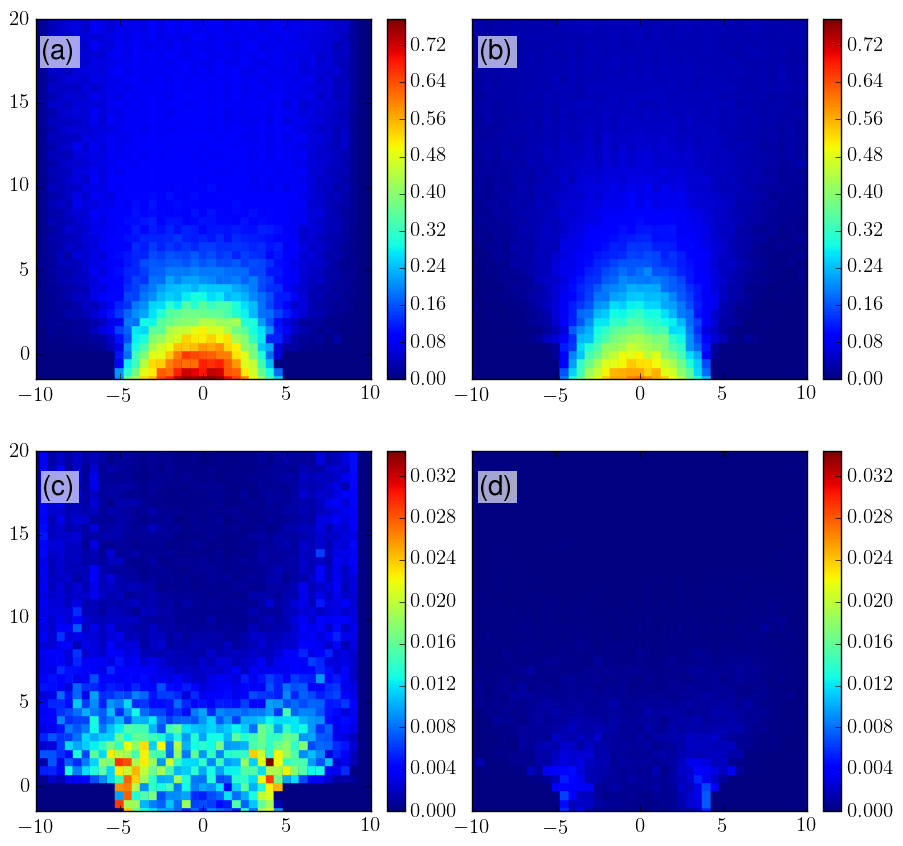}
\caption{Mean kinetic energy of the particles: (a) translational for disks, (b) translational for triangles, (c) rotational for disks and (d) rotational for triangles. The aperture is $D=10d$ and the friction coefficient $\mu=0.5$. Disks and triangles have the same area (mass). The color scale indicates the energy in J.}
\label{fig6}
\end{figure}

In Fig. \ref{fig6} we show the translational $k_t$ and rotational $k_r$ kinetic energy profile for disks and triangles. $k_t$ is dominated by the vertical component of the velocity and therefore the features observed in Fig. \ref{fig5}(c) and Fig. \ref{fig5}(d) are reproduced. An interesting feature observed in $k_r$ is that disks present a wide region over which grains rotate. This region extends all along the orifice width (and beyond, over the flat bottom base) at a height about $2d$ above the aperture. Triangles, in marked contrast, only acquire rotational energies less than half those of disks and over a limited region next to each edge of the aperture. This might be due to the frustration of any initial spin a particle may acquire since collisions of its sharp edges with neighboring grains will tend to stop any rotational motion.

At first sight, the reduced rotational energy of triangles seems counter-intuitive. If triangles have a reduced translational energy, one should expect that part of the energy input provided by the work done by the force of gravity might go into the rotational degrees of freedom. Notice however that the energy is also dissipated by collisions and friction. It is clear then that triangles provide a much efficient dissipation of the energy for all degrees of freedom.

\subsection{The effective size of a grain}

It is important to notice that comparing the behavior of grains of different shapes conveys an arbitrary decision on the parameter that decides when two grains are considered of the same size. In the previous section we have chosen to define as particles of the same size those that have the same mass, irrespective of the shape. However, in some applications, one may be interested in comparing the flow of different polygons that have the same radius, side or perimeter.

\begin{figure}
\includegraphics[width=1.0\columnwidth]{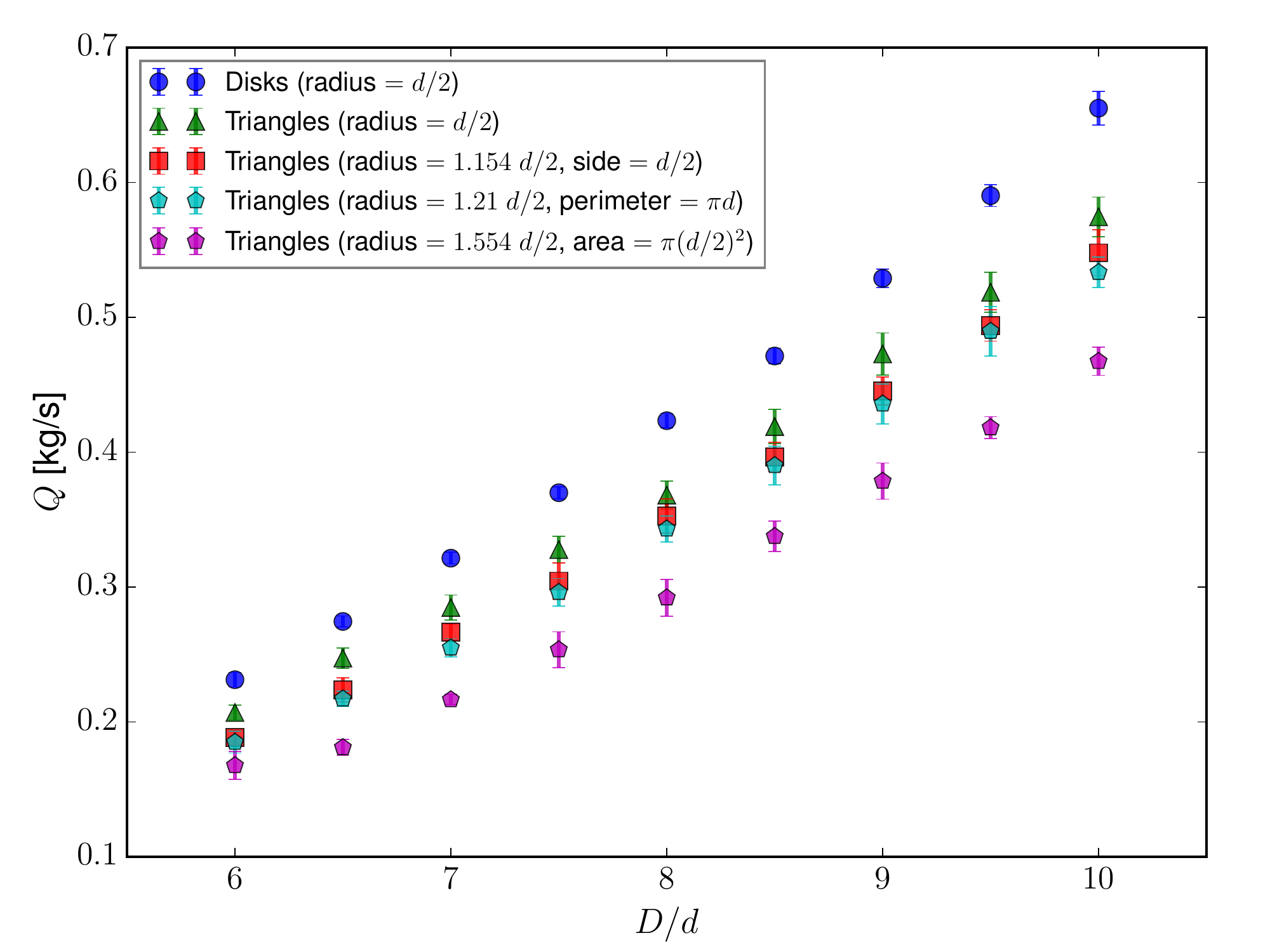}
\caption{Mass flow rate as a function of the aperture size $D$ for disks and triangles with $\mu=0.5$. Different triangles, that match the diameter, perimeter and area of the disks as well as having the side equal to the diameter of a disk, are considered.}
\label{fig7}
\end{figure}

Figure \ref{fig7} shows the mass flow rate for triangular particles of different sizes chosen to match the radius, perimeter or area of the reference disks. We have also included data for triangles whose side is the same as the diameter of the disks. As we can see, since the actual radius of the triangle depends on the parameter used to define if the polygon is ``of the same size as the disk'', the mass flow rate varies up to 25\%. 

The importance of the arbitrary definition of the effective size of the grains when comparing different shapes was previously recognized \cite{hohner2013}. However, the extent of the effect that this may cause in the mass flow rate was not studied. The 25\% maximum difference observed in $Q$ for triangles show that this cannot be neglected, although other shapes may be less affected. 

Notice that it is usual to compare particles of different shape that have the same mass. However, there exist several works where authors use a characteristic length instead of the mass \cite{cleary2002,hohner2012,hohner2013}. In some applications, due to design, economic cost, practical constraints, etc., fabricated objects of different shape may or may not have the same mass. Estimations of the relative mass flow rate of such objects delivered by hoppers in comparison with circular counterparts would need to take into account the relative size. 

\subsection{Particle flow rate}

\begin{figure}
\includegraphics[width=0.9\columnwidth]{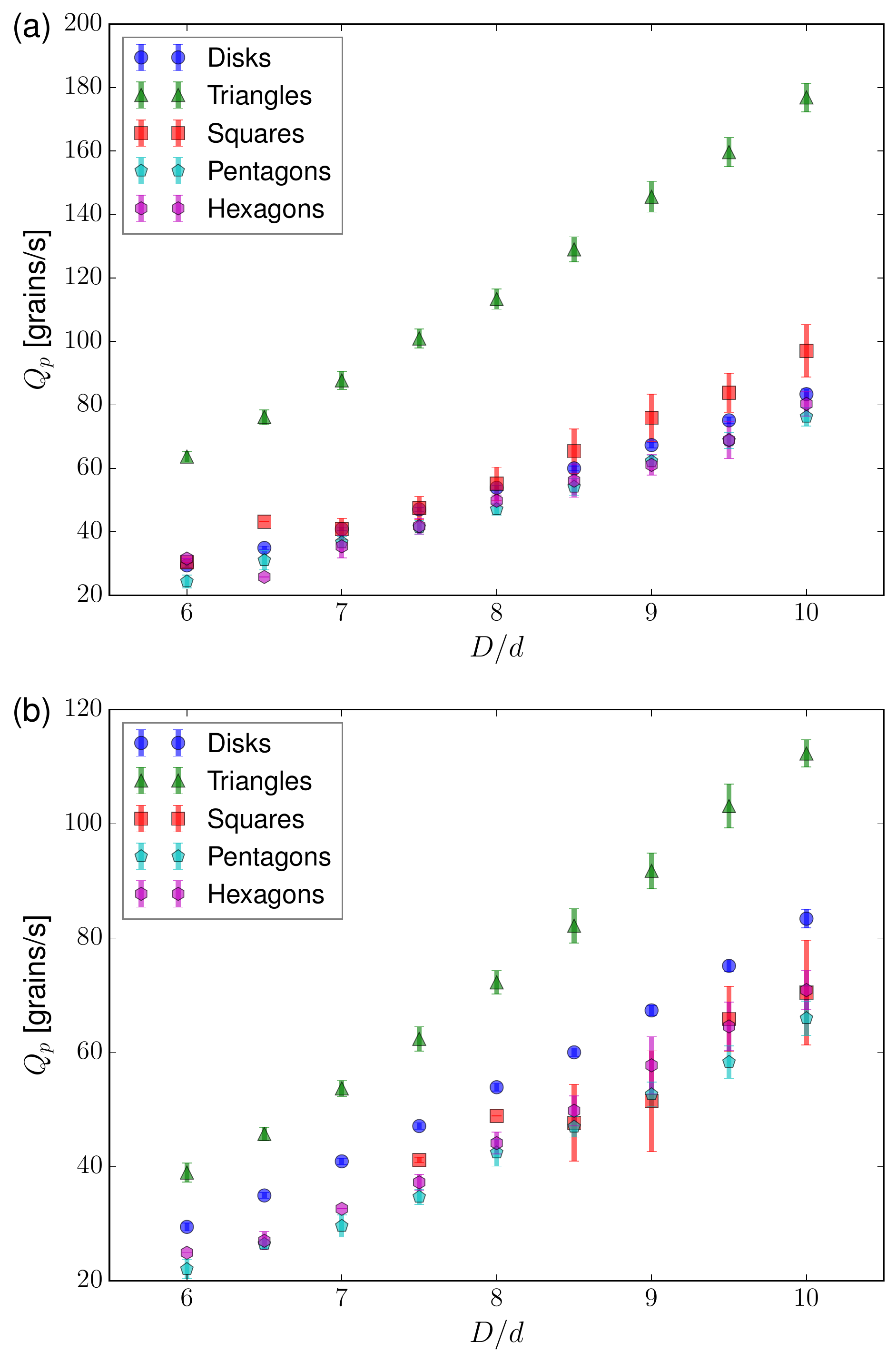}
\caption{(a) Particle flow rate as a function of the aperture size $D$ for different particle shapes with $\mu=0.5$ and the same particle radius. (b) Particle flow rate as a function of the aperture size $D$ for different particle shapes with $\mu=0.5$ and the same particle perimeter.}
\label{fig8}
\end{figure}

It is worth mentioning that in some cases, the number of grains flowed per unit time may be the important parameter rather than the mass flow rate. Although this is proportional to the mass flow rate, the proportionality constant depends on the size of the grains and the mean $\phi$. Hence, depending on whether polygons are compared by their radius, side or perimeter, the actual particle flow rate, $Q_p$ varies. Naturally, if polygonal grains are compared by their area in 2D, the particle flow rate yields the same results as in Fig. \ref{fig2} simply scaled by the same factor (the particle density) for all shapes. However, this is not the case if grains have, for example, the same diameter. 

In Fig. \ref{fig8} we present $Q_p$ for different polygons if the radius (a) or the perimeter (b) is used as the reference to compare sizes. As we can see, the relative position of the curves changes significantly depending on the parameter used to compare the particle sizes since the relative masses of the grains are different. If particles have the same radius [Fig. \ref{fig8}(a)], pentagons and hexagons yield a slightly lower particle flow rate than disks, while squares deliver a larger number of particles per unit time. Triangles present up to 100\% higher particle flow rates than disks. This is in dramatic contrast with the mass flow rate observed in Fig. \ref{fig7}. If we use the perimeter to compare grain sizes, most polygons have a lower particle flow rate than disks while triangles still display a larger flow in number of grains per seconds. 

The results of Fig. \ref{fig8} show that the particle flow rate is particularly sensitive to the effective size of the grains. Squares, for example, may be delivered at a higher or lower rate than disks,  in terms of number of particles, depending on whether they have the same radius or the same perimeter as disks. However, the mass flow rate of squares is lower than that of disks in both cases.

\section{Conclusions}
\label{sec:3}

We have considered the discharge of circular and polygonal grains from a flat bottomed silo. We have seen, in agreement with other workers, that for grains of the same mass, the mass flow rate is markedly lower for polygons in comparison with disks. We have found that this is caused by a combination of reduced mean packing fraction for the flowing polygons and also a reduced mean vertical velocity. The rotational kinetic energy of the polygons is also significantly smaller than that of disks.

When comparing results for different shapes, it is necessary to arbitrarily select a parameter to define when two grains can be said to be of equal size. We found that this selection, that may be constrained by specific applications, may lead to significant changes in the actual mass flow rates.

We have discussed that the ratio of the particle flow rate to the mass flow rate is a non-trivial function of the grain shape since it does not only depend on the mass of the grains but also on the actual packing fraction. As a consequence, it is difficult to predict the particle flow rate from the sole knowledge of the mass flow rate. We have shown that the particle flow rate for a given shape, which can be the measure of interest in many industrial applications, can result higher or lower than for a different particle shape depending on the parameter used to define their effective size. Hence, in a specific application, in order to deliver the same number of grains per unit time for a different particle shape, the aperture must be either reduced or widened depending on the parameter used to compare their sizes (e.g., diameter, mass, perimeter, etc).

\begin{acknowledgements}
This work has been supported by ANPCyT (Argentina) through grant PICT 2012-2155 and UTN (Argentina) through grants PID MAUTNLP0002184 and PID IFI1871.
\end{acknowledgements}


\end{document}